\documentclass{./styles/svproc}
\usepackage{url}
\usepackage{graphicx}
\usepackage{float}

\usepackage[margin=1in]{geometry}
\usepackage{amsmath}
\usepackage{amssymb}
\usepackage{booktabs}
\usepackage{multirow}
\usepackage{algorithm}
\usepackage{algorithmic}
\usepackage{authblk}
\usepackage{subcaption}
\usepackage{tcolorbox}

\usepackage{bbm}

\usepackage{amsmath}                 
\usepackage{amssymb}                 
\usepackage{amsfonts}                
\usepackage{mathtools}               
\usepackage{bm}                      

\usepackage{booktabs}                
\usepackage{array}                   
\usepackage{multirow}                
\usepackage{makecell}                
\usepackage{tabularx}                

\usepackage{graphicx}                
\usepackage{float}                   
\usepackage{subcaption}              
\usepackage[dvipsnames]{xcolor}      

\usepackage{tikz}                    
\usetikzlibrary{
    trees,                           
    positioning,                     
    shapes.geometric,                
    arrows.meta,                     
    calc,                            
    fit,                             
    backgrounds                      
}

\usepackage[
    font=small,
    labelfont=bf,
    labelsep=period
]{caption}                           

\usepackage{enumitem}                
\setlist{nosep}                      

\usepackage[
    colorlinks=true,
    linkcolor=blue!70!black,
    citecolor=green!50!black,
    urlcolor=blue!70!black
]{hyperref}                          
\usepackage{url}                     

\usepackage{pifont}                  

\usepackage{algorithm}               
\usepackage{algorithmic}             

\usepackage{listings}                
\usepackage{xspace}                  
\usepackage{lipsum}                  

\begin{document}
\mainmatter              

\title{LLM-Mediated Demand Response Coordination in Smart Microgrids}

\titlerunning{LLM-Mediated Demand Response Coordination in Smart Microgrids}

\author{J. de Curt\`o\inst{1,2} and I. de Zarz\`a\inst{3}}

\authorrunning{de Curt\`o et al.}

\tocauthor{de Curt\`o et al.}

\institute{
Department of Computer Applications in Science \& Engineering, BARCELONA Supercomputing Center, 08034 Barcelona, Spain 
\and 
Escuela Técnica Superior de Ingeniería (ICAI), Universidad Pontificia Comillas, 28015 Madrid, Spain
\and
Human Centered AI, Data \& Software, LUXEMBOURG Institute of Science and Technology, L-4362 Esch-sur-Alzette, Luxembourg 
}

\maketitle              

\begin{abstract}

Effective demand response in smart microgrids requires prosumers to cooperate 
voluntarily under conditions of strategic self-interest, a coordination problem 
structurally equivalent to a repeated Prisoner's Dilemma on a social network. 
This paper presents a multi-agent simulation in which a Large Language Model (LLM) Influence Compiler 
issues structured demand-response directives to a population of heterogeneous 
prosumer agents, each governed by a hybrid decision architecture combining 
game-theoretic base probability (derived from payoff history, neighbour imitation, 
and exploitation memory) with LLM narrative evaluation of incoming coordination 
signals. The hybrid architecture resolves a critical methodological challenge: 
LLMs aligned via Reinforcement Learning from Human Feedback (RLHF) exhibit overwhelming cooperation bias when used as 
direct decision-makers, producing flat dynamics regardless of grid conditions. By 
separating strategic reasoning from grounded narrative evaluation, the model 
generates realistic prosumer behaviour across six personality archetypes, pragmatist, 
idealist, skeptic, conformist, strategist, and opportunist, with baseline cooperation 
near 50\% and meaningful differentiation under influence. Compiled structured 
directives achieve 33.3\% demand-curtailment cooperation versus 27.0\% for 
unstructured messaging and 28.0\% for a no-intervention baseline 
($\Delta_{\mathrm{comp}} = +0.063$), with the advantage preserved across both 
grounded and idealized agent substrates ($\Delta = +0.083$) and across all 
resistance levels ($R = 0.1$--$0.7$). Hub-targeted dissemination via 
high-centrality network nodes outperforms peripheral or random targeting, 
confirming that grid topology provides mechanistic amplification independent of 
message content. These results suggest that structured LLM compilation, grounded 
agent reasoning, and network-aware targeting are complementary design principles 
for scalable, interpretable demand-response coordination in next-generation 
smart city energy systems.

\keywords{demand response coordination, smart microgrids, large language models, 
multi-agent systems, evolutionary game theory}
\end{abstract}

\section{Introduction}
\label{sn:intro}

The transition toward smart city energy infrastructures has placed voluntary demand 
response coordination at the centre of grid resilience research~\cite{szabo2007evolutionary,santos2005scale,deZarza2025}. In next-generation 
microgrids~\cite{fang2011smart,siano2014demand}, prosumers, residential and commercial agents who both consume and 
produce electricity, must decide whether to curtail their demand during peak load 
events or to free-ride on the curtailment of others. This constitutes a coordination 
problem structurally equivalent to a repeated Prisoner's Dilemma (PD) on a social 
network~\cite{axelrod1984evolution,nowak2006five}: collective welfare is maximised 
when all prosumers cooperate, yet individual incentives favour defection. Classical 
demand-response mechanisms rely on price signals or contractual obligations, but 
increasingly, smart grid operators seek communication-based coordination strategies 
that can adapt in real time to heterogeneous prosumer populations without requiring 
coercive enforcement~\cite{szabo2007evolutionary}.

Large language models (LLMs) have recently demonstrated the capacity to act as 
deliberative agents in multi-agent systems, generating advice, mediating negotiation, 
and influencing the strategic behaviour of interacting 
parties~\cite{park2023generative,guo2024large,xi2023rise}. Their ability to produce 
contextualised natural-language narratives makes them attractive candidates for 
issuing demand-response directives that are interpretable to human prosumers and 
adaptable to their individual circumstances. Several studies have examined 
LLM behaviour in canonical game-theoretic settings~\cite{akata2023playing,fan2024can,fontana2024nicer}, 
consistently finding that RLHF-aligned models exhibit a strong prosocial bias that 
can distort experimental results when LLMs serve simultaneously as influence 
generators and as agent decision-makers. This cooperative bias produces 
near-uniform compliance regardless of incentive structure, obscuring the treatment 
effects that coordination research seeks to measure.

Prior work established that theory-grounded coevolution combined with LLM 
deliberation outperforms either mechanism in isolation for achieving stable 
cooperation in networked multi-agent populations~\cite{deZarza2023,deCurto2025}. 
This paper addresses the limitations of the previous studies by introducing a \emph{hybrid 
decision architecture} that separates game-theoretic reasoning from LLM narrative 
evaluation. In the proposed model, a game-theoretic base probability, computed 
from payoff history, exploitation memory, and neighbour imitation following 
Fermi-function social learning~\cite{szabo1998evolutionary,traulsen2007pairwise}, 
creates genuine strategic tension with baseline cooperation near $50\%$. An LLM 
narrative evaluation layer then assesses incoming coordination signals and 
contributes a cooperation shift bounded in $[-0.30, +0.30]$, attenuated by 
agent-specific resistance. This hybrid architecture preserves the core research 
question, whether structured compilation improves coordination signal 
effectiveness, while preventing RLHF cooperation bias from collapsing all 
experimental conditions to identical dynamics~\cite{piatti2024cooperate}.

We instantiate this framework in a smart microgrid demand-response scenario. 
Six prosumer personality archetypes, pragmatist, idealist, skeptic, conformist, 
strategist, and opportunist, are distributed over a scale-free network 
topology~\cite{barabasi1999emergence}, reflecting the heterogeneous population 
typical of urban energy communities. A central Influence Compiler observes a noisy 
snapshot of the population state and issues structured demand-curtailment directives 
every five time steps to $20\%$ of agents, with non-targeted prosumers receiving 
attenuated word-of-mouth signals from their network neighbours.

Our experiments address four research questions. \textbf{(RQ1)} Does compiled 
influence lift demand-curtailment cooperation above both the no-intervention 
baseline and the unstructured messaging baseline? \textbf{(RQ2)} How does the 
compilation advantage compare between idealized and grounded agent architectures? 
\textbf{(RQ3)} Does hub-targeted dissemination preserve its advantage over random 
and peripheral targeting when agents can resist? \textbf{(RQ4)} How does the compilation advantage degrade as prosumer 
resistance increases, and does it vanish within the tested range?

\section{Related Work}
\label{sn:related}

The deployment of large language models as autonomous agents in multi-agent 
environments has accelerated significantly following the demonstration that 
LLM-powered agents can sustain coherent persona, episodic memory, and social 
interaction over extended simulation horizons~\cite{park2023generative}. 
Comprehensive surveys of this landscape identify role specialisation, 
inter-agent communication, and collective decision-making as the central open 
problems~\cite{guo2024large,xi2023rise}. Architectures based on multi-agent 
debate have shown that assigning adversarial roles to distinct LLM instances 
improves factual accuracy and divergent reasoning beyond what single-model 
prompting achieves~\cite{du2024improving,liang2024encouraging}, while the CAMEL 
framework demonstrated that communicative role-playing agents can solve complex 
tasks through structured dialogue without human 
intervention~\cite{li2024camel}.

The specific question of whether LLMs can sustain cooperative equilibria in 
strategic settings has attracted growing attention. \cite{akata2023playing} 
showed that GPT-class models playing repeated games exhibit higher cooperation 
rates than human baselines, a finding echoed by \cite{fontana2024nicer}, who 
found that LLMs are systematically more cooperative than humans in Prisoner's 
Dilemma experiments. \cite{fan2024can} provided a systematic analysis of LLM 
rationality across canonical game-theoretic scenarios, identifying consistent 
deviations from NASH equilibrium play. Crucially, \cite{piatti2024cooperate} 
demonstrated that cooperation among LLM agents is fragile: it collapses under 
resource scarcity and adversarial defection unless stabilised by structural 
mechanisms. Our work extends this line by showing that the RLHF-induced 
cooperation bias documented in these studies renders LLMs unsuitable as 
\emph{direct} decision-makers in controlled cooperation experiments, motivating 
the hybrid architecture introduced in Section~\ref{sn:model}.

The emergence of cooperation in populations of self-interested agents is one of 
the central problems of evolutionary game theory~\cite{axelrod1984evolution}. 
\cite{nowak2006five} identified five mechanisms, kin selection, direct 
reciprocity, indirect reciprocity, network reciprocity, and group 
selection, through which natural selection can favour cooperative strategies. 
Network reciprocity, in which spatial or social structure clusters cooperators 
and shields them from exploitation by defectors, is particularly relevant to 
prosumer populations connected through energy community 
networks~\cite{santos2005scale,ohtsuki2006simple,newman2003structure}.

Scale-free networks, characterised by a power-law degree 
distribution~\cite{barabasi1999emergence}, have been shown to provide an 
especially strong substrate for cooperation because hub nodes accumulate payoff 
advantages that diffuse through the network via Fermi-function imitation 
dynamics~\cite{szabo1998evolutionary,traulsen2007pairwise}. \cite{santos2006cooperation} 
demonstrated that allowing agents to rewire their social ties further amplifies 
this effect. The coevolutionary dynamics studied by \cite{perc2010coevolutionary} 
and the statistical physics perspective of \cite{perc2017statistical} provide the 
theoretical backbone for the EC-theory update rule employed in our simulation. 
Small-world topology~\cite{watts1998collective} and Erd\H{o}s--R\'enyi random 
graphs~\cite{erdos1959random} serve as comparative baselines, with scale-free 
topology as the primary setting given its correspondence to real-world energy 
community structures where a small number of prosumers or aggregators hold 
disproportionate network influence.

Demand response, the voluntary adjustment of electricity consumption by end users 
in response to grid signals, is a cornerstone of smart microgrid 
flexibility~\cite{siano2014demand,parag2016electricity}. Traditional approaches rely on 
time-of-use pricing, direct load control contracts, or automated building 
management systems. The integration of artificial intelligence into grid 
management~\cite{khan2022artificial,madani2025large,shi2024review}, explicitly envisioned in the smart grid and microgrid literature 
under the rubric of layered distributed intelligence, opens the possibility of 
communication-based coordination that adapts to prosumer heterogeneity without 
coercive enforcement.

The prosumer dilemma, whether to curtail demand for collective grid 
stability or to free-ride on others' curtailment, is structurally equivalent 
to the public goods game and the Prisoner's Dilemma, both extensively studied in 
evolutionary game theory~\cite{hauert2004spatial,traulsen2006stochastic}. Game-theoretic 
models of demand response have shown that peer effects and social network 
structure significantly influence curtailment decisions, motivating 
network-aware targeting strategies. Our contribution bridges the game-theoretic 
demand response literature and the emerging LLM multi-agent literature by 
instantiating the Influence Compiler framework in an energy coordination 
scenario, with hybrid prosumer agents whose strategic behaviour is grounded in 
both payoff history and LLM narrative evaluation of coordination signals.

The risk that a capable influence compiler could achieve coordination 
through manipulative or inequitable messaging, for instance, fear-based 
framing or biased subgroup targeting, motivates future governance extensions 
to the present framework. Constitutional AI approaches~\cite{anthropic2024claude,huang2024collective} 
provide a natural template for filtering such directives through explicit 
normative constraints, and we identify this as a primary direction for 
subsequent work.

\section{Model and Simulation Setup}
\label{sn:model}

We model a smart microgrid energy community as a population of $N = 30$ prosumer 
agents distributed over a scale-free network $\mathcal{G} = (\mathcal{V}, 
\mathcal{E})$ generated via the Barab\'asi--Albert preferential attachment 
mechanism with $m = 3$~\cite{barabasi1999emergence}. At each discrete time step 
$t \in \{1, \ldots, T\}$ with $T = 50$, every agent $o \in \mathcal{V}$ makes a 
binary demand-curtailment decision $a_o(t) \in \{C, D\}$, where $C$ denotes 
cooperative curtailment and $D$ denotes non-cooperative consumption. The payoff 
structure follows a standard Prisoner's Dilemma with $R = 3$ (mutual 
curtailment), $T = 5$ (unilateral non-curtailment), $S = 0$ (unilateral 
curtailment while neighbour consumes), and $P = 1$ (mutual non-curtailment), 
satisfying $T > R > P > S$ and $2R > T + S$.

A central Influence Compiler $\mathcal{C}$ observes a noisy snapshot of the 
population state every $\Delta t = 5$ steps and deploys a compiled 
demand-response directive to a targeted subset of $\lfloor 0.2N \rfloor = 6$ 
agents. Non-targeted agents receive attenuated word-of-mouth signals from 
cooperating neighbours with probability proportional to the local cooperation 
rate in their neighbourhood. The Influence Compiler $\mathcal{C}$ abstracts the role of a demand-response 
aggregator or distribution system operator, which in real microgrid deployments 
issues coordination signals to enrolled prosumers~\cite{siano2014demand,parag2016electricity}.

\subsection{Hybrid Decision Architecture}

The central methodological contribution of this work is the \emph{hybrid decision 
architecture} (v3), which decouples game-theoretic base probability from LLM 
narrative evaluation to prevent RLHF cooperation bias from collapsing experimental 
conditions~\cite{piatti2024cooperate,fontana2024nicer}. The cooperation 
probability for agent $o$ at time $t$ is:

\begin{equation}
  p_o^C(t) \;=\; \mathrm{clip}\!\left(\,p_o^{\mathrm{base}}(t) \;+\; 
  \delta_o^{\mathrm{LLM}}(t)\cdot\bigl(1 - r_o\bigr),\; 0.02,\; 0.98\right),
  \label{e:hybrid}
\end{equation}

\noindent where $p_o^{\mathrm{base}}(t)$ is the game-theoretic base probability, 
$\delta_o^{\mathrm{LLM}}(t) \in [-0.30, +0.30]$ is the narrative evaluation 
shift produced by the LLM, and $r_o \in [0,1]$ is the agent's fixed resistance 
level. The clipping ensures a non-trivial probability of either action for all 
agents at all times.

\subsubsection{Stage 1: Game-Theoretic Base Probability}

The base probability $p_o^{\mathrm{base}}(t)$ aggregates five behavioural 
mechanisms grounded in evolutionary cooperation theory~\cite{perc2017statistical,traulsen2006stochastic}:

\begin{equation}
  p_o^{\mathrm{base}}(t) \;=\; \sigma\!\left(\,w_1\,\phi_o^{\mathrm{arch}} 
  \;-\; w_2\,e_o(t) \;+\; w_3\,\bar{c}_{\mathcal{N}(o)}(t) 
  \;-\; w_4\,\Delta\pi_o(t) \;-\; w_5\,\tau_o(t)\right),
  \label{e:base}
\end{equation}

\noindent where $\sigma(\cdot)$ is the logistic function; $\phi_o^{\mathrm{arch}}$ 
is an archetype-specific base bias; $e_o(t)$ is the normalised exploitation 
count (proportion of past rounds in which agent $o$ cooperated while the majority 
of neighbours defected); $\bar{c}_{\mathcal{N}(o)}(t)$ is the mean cooperation 
rate of $o$'s neighbourhood (conformist imitation~\cite{szabo1998evolutionary}); 
$\Delta\pi_o(t) = \pi_o^D(t) - \pi_o^C(t)$ is the payoff advantage of defection 
over cooperation in the most recent round; and $\tau_o(t)$ is a temporal 
temptation decay term that increases defection probability when consecutive 
defections by neighbours go unpunished. The weights $\{w_k\}$ are set to $\{0.5, 1.2, 0.8, 0.6, 0.3\}$, calibrated to 
produce baseline cooperation near $50\%$ without influence. This creates genuine 
strategic tension across heterogeneous agents.

\subsubsection{Stage 2: LLM Narrative Evaluation}

When agent $o$ is targeted by the Influence Compiler, it receives a 
demand-response directive as natural language. Rather than asking the LLM to 
make a binary curtailment decision (which produces near-universal cooperation 
due to RLHF alignment), the agent is prompted under a 
\texttt{NARRATIVE\_EVAL\_SYSTEM} instruction that asks: \emph{given your 
persona, memory, and this message, by how much does your willingness to curtail 
demand change?} The LLM outputs a scalar shift $\delta_o^{\mathrm{LLM}} \in 
[-0.30, +0.30]$, where positive values indicate increased curtailment 
willingness and negative values indicate backlash. The prompt encodes the 
agent's personality archetype, the last five rounds of payoff history, the 
most recent neighbourhood cooperation rate, and the full text of the compiled 
directive.

Non-targeted agents receive no LLM call; their cooperation probability is 
determined entirely by $p_o^{\mathrm{base}}(t)$ plus any word-of-mouth shift 
drawn from $\mathcal{N}(0, 0.05)$ if at least one cooperating neighbour 
transmitted the directive.

\subsubsection{Stage 3: Resistance Attenuation and Sampling}

The effective shift is attenuated by agent resistance: 
$\delta_o^{\mathrm{eff}}(t) = \delta_o^{\mathrm{LLM}}(t) \cdot (1 - r_o)$, 
so that an agent with $r_o = 0.7$ retains only $30\%$ of any narrative 
influence. The binary action is then sampled from a Bernoulli distribution with parameter 
$p_o^C(t)$ as defined in Equation~\ref{e:hybrid}.

\subsection{Personality Archetypes}

Six prosumer archetypes encode heterogeneous attitudes toward demand curtailment, 
reflecting the diversity of real urban energy community 
participants~\cite{perc2010coevolutionary}:

\begin{itemize}
  \item \textbf{Pragmatist} ($\phi = 0.00$, $r = 0.20$): payoff-driven, 
        responds proportionally to demonstrated grid benefit.
  \item \textbf{Idealist} ($\phi = +0.10$, $r = 0.15$): intrinsically 
        motivated toward collective welfare; most responsive to moral framing.
  \item \textbf{Skeptic} ($\phi = 0.00$, $r = 0.55$): requires strong 
        evidence before changing behaviour; resistant to all framing themes.
  \item \textbf{Conformist} ($\phi = 0.00$, $r = 0.20$): follows neighbourhood 
        majority; most influenced by social norm signals.
  \item \textbf{Strategist} ($\phi = -0.05$, $r = 0.30$): punishes defectors 
        via conditional cooperation; responds to reciprocity-based framing.
  \item \textbf{Opportunist} ($\phi = -0.15$, $r = 0.45$): free-rider tendency; 
        exploits cooperative neighbours and resists curtailment directives.
\end{itemize}

\noindent Archetypes are assigned to agents uniformly at random. Each agent 
maintains an episodic memory buffer of the last ten rounds, tracking cooperation 
decisions, received payoffs, and exploitation events (rounds in which the agent 
curtailed while the majority of neighbours consumed).

\subsection{LLM Influence Compiler}

The Influence Compiler $\mathcal{C}$ operates on a noisy state snapshot 
$\hat{s}(t)$, in which prosociality and cooperation estimates are perturbed by 
independent Gaussian noise $\mathcal{N}(0, 0.10)$ to model imperfect population 
observability. Given $\hat{s}(t)$, the compiler generates a structured policy:

\begin{equation}
\begin{aligned}
  \pi \;=\; \Bigl(\;&\texttt{target},\;\;
    \texttt{intensity} \in [0,1],\;\;
    \texttt{timing} \in \{\texttt{burst},\,\texttt{periodic}\},\\
    &\texttt{theme} \in \{\texttt{moral},\,\texttt{economic},\,
    \texttt{identity},\,\texttt{hybrid}\}\;\Bigr),
\end{aligned}
  \label{e:policy}
\end{equation}

\noindent and then renders $\pi$ into a natural-language directive through a 
Solver--Critic pipeline~\cite{deCurto2025}: the Solver proposes a policy and 
drafts the directive; the Critic evaluates alignment with the policy parameters 
and may request revision. The final directive is delivered to the targeted 
agent subset. In the \emph{unstructured} baseline, the compiler is replaced by 
an unconstrained free-form narrative generator with no policy schema.

Targeting strategies follow four conditions: \texttt{HUBS} (top-$k$ agents by 
degree centrality), \texttt{BRIDGES} (top-$k$ agents by betweenness centrality), 
\texttt{PERIPHERY} (bottom-$k$ agents by degree), and \texttt{RANDOM} (uniform 
sampling). All LLM calls use Llama-3.3-70B-Instruct via the Nebius AI Studio 
API~\cite{touvron2023llama,grattafiori2024llama} with SHA1-based caching for reproducibility.

\subsection{Evaluation Metrics}

The primary outcome metric is the \emph{population cooperation rate} 
$\bar{c}(t) = \frac{1}{N}\sum_{o=1}^{N} \mathbf{1}[a_o(t) = C]$, averaged 
over the final ten time steps to reduce stochastic noise. Secondary metrics 
include: \emph{final strategy} $s_o \in [0,1]$ (agent-level cooperation 
propensity estimated from the last ten rounds); \emph{persuasion rate} 
$\rho_o$ (proportion of influence deployments that produced 
$\delta_o^{\mathrm{LLM}} > 0$); and \emph{backlash rate} $\beta$ (proportion 
of deployments producing $\delta_o^{\mathrm{LLM}} < 0$). The 
\emph{compilation advantage} is defined as 
$\Delta_{\mathrm{comp}} = \bar{c}_{\mathrm{compiled}} - 
\bar{c}_{\mathrm{unstructured}}$.

\section{Experimental Results}
\label{sn:experiments}

We report four experiments addressing the research questions stated in 
Section~\ref{sn:intro}. All cooperation rates are averaged over the final ten 
time steps ($t \in [40, 49]$) to reduce stochastic noise. The full time-series 
data for each condition are available in the supplementary material.

\subsection{Experiment 1: Compiled vs.\ Unstructured vs.\ No-Influence 
(RQ1)}
\label{subsn:exp1}

The first experiment compares three conditions on a population of $N = 30$ 
grounded hybrid agents over $T = 50$ time steps: (1)~compiled structured 
directives issued by the Influence Compiler, (2)~unstructured free-form 
narrative generation without a policy schema, and (3)~no-influence baseline 
in which game-theoretic dynamics operate without any external intervention.

\begin{figure}[ht]
  \centering
  \includegraphics[width=0.8\linewidth]{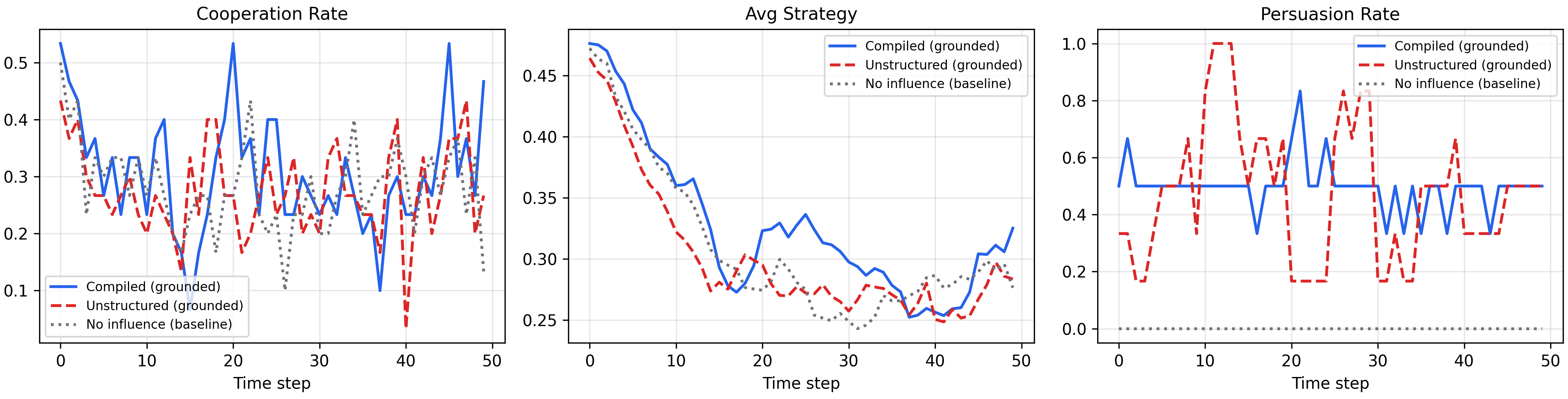}
\caption{Experiment~1: Compiled versus unstructured versus no-influence 
           conditions with grounded hybrid agents. Left: cooperation rate 
           time series. Centre: average strategy trajectory. Right: 
           persuasion rate over time.}
  \label{fgr:exp1}
\end{figure}

Figure~\ref{fgr:exp1} shows the cooperation rate time series, average strategy 
trajectory, and persuasion rate over time. The compiled condition achieves a 
final cooperation rate of $0.333$ and a time-averaged rate of $0.305$, compared 
to $0.270$ (final) and $0.275$ (average) for unstructured messaging. The 
no-influence baseline settles at $0.280$ (final), $0.282$ (average), confirming 
that unstructured messaging performs \emph{below} the no-intervention level. 
The compilation advantage is $\Delta_{\mathrm{comp}} = +0.063$ in final 
cooperation rate, and the lift above baseline is $+0.053$, confirming 
\textbf{RQ1}: compiled influence meaningfully raises demand-curtailment 
cooperation while unstructured messaging produces a slight backfire effect.

The average strategy trajectory (Figure~\ref{fgr:exp1}, centre) shows all three 
conditions initially decaying from $\sim\!0.47$ as game-theoretic incentives 
erode prosocial priors; the compiled condition then recovers to $0.325$, while 
unstructured ($0.284$) and baseline ($0.276$) continue declining. Persuasion 
rates (right) fluctuate around $0.50$ for both influence conditions, confirming 
that the cooperation gap is driven by the quality of $\delta_o^{\mathrm{LLM}}$, 
not its frequency.

\subsection{Experiment 2: Grounded vs.\ Idealized Agent Substrate (RQ2)}
\label{subsn:exp2}

The second experiment quantifies the cost of agent realism by comparing 
grounded hybrid agents (Experiment~1) against idealized logistic responders 
under otherwise identical conditions. Table~\ref{t:exp2} summarises the 
results across all three conditions for both agent substrates.

\begin{table}[t]
\centering
\caption{Experiment~2: Final cooperation rates under grounded and idealized 
         agent substrates.}
\label{t:exp2}
\begin{tabular}{lcccc}
\hline
\textbf{Condition} & \textbf{Grounded} & \textbf{Idealized} & 
\textbf{$\Delta$ (G)} & \textbf{$\Delta$ (I)} \\
\hline
Compiled     & 0.333 & 0.850 & \multirow{2}{*}{+0.063} & 
\multirow{2}{*}{+0.083} \\
Unstructured & 0.270 & 0.767 & & \\
No influence & 0.280 & 0.723 & --- & --- \\
\hline
\end{tabular}
\end{table}

\begin{figure}[H]
  \centering
  \includegraphics[width=0.6\linewidth]{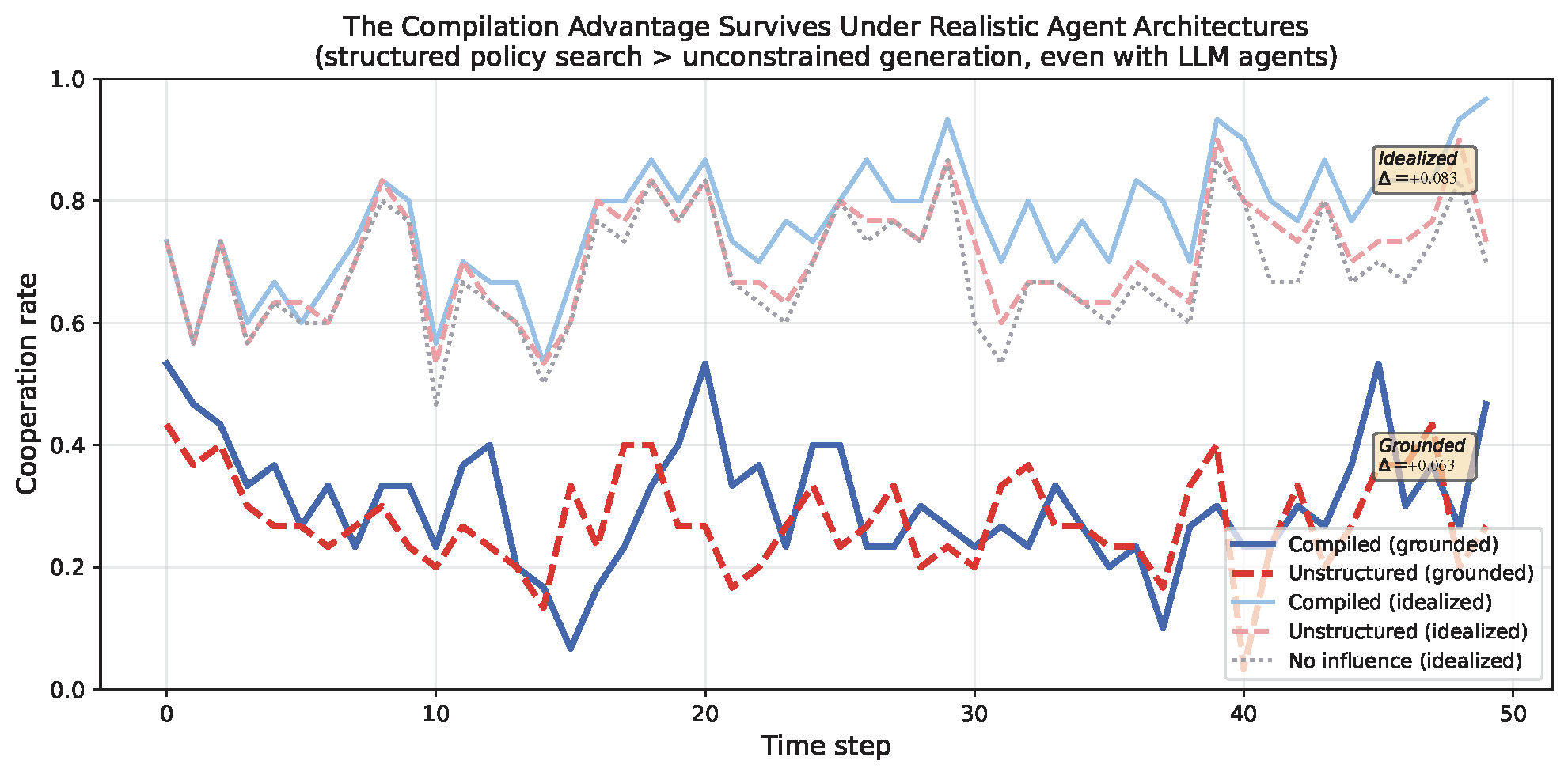}
\caption{Compiled (blue) versus unstructured (red dashed) influence under 
          idealized and grounded agent substrates.}
  \label{fgr:key_message}
\end{figure}

Idealized agents reach $0.850$ cooperation under compiled influence versus 
$0.767$ under unstructured, yielding an idealized compilation advantage of 
$+0.083$. The grounded compilation advantage ($+0.063$) is somewhat smaller, 
reflecting the $-0.517$ absolute cooperation gap introduced by genuine strategic 
resistance, exploitation memory, and personality heterogeneity. Critically, 
the \emph{relative ordering} of conditions is fully preserved: compiled 
$>$ unstructured $>$ no-influence holds under both substrates, and the 
direction and significance of the compilation advantage is robust. This 
confirms \textbf{RQ2}: the compilation advantage is a genuine compiler design 
property that survives the transition from idealized to realistic agent 
architectures, as also visualised in Figure~\ref{fgr:key_message}.

\subsection{Experiment 3: Targeting Robustness (RQ3)}
\label{subsn:exp3}

The third experiment evaluates four targeting strategies under compiled 
influence with grounded agents: hub-targeting (\texttt{HUBS}), bridge-targeting 
(\texttt{BRIDGES}), random targeting (\texttt{RANDOM}), and periphery-targeting 
(\texttt{PERIPHERY}).

\begin{figure}[H]
  \centering
  \includegraphics[width=0.8\linewidth]{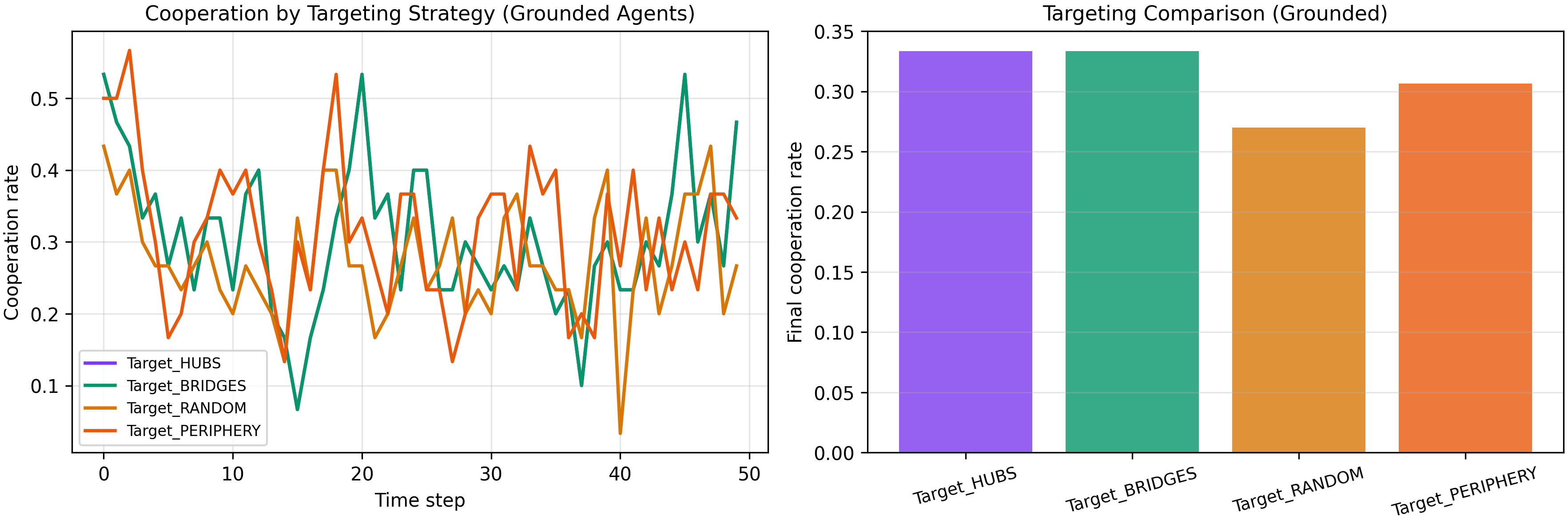}
\caption{Experiment~3: Cooperation rate by targeting strategy over time 
           (left) and final cooperation rate comparison (right).}
  \label{fgr:exp3}
\end{figure}

Figure~\ref{fgr:exp3} and Table~\ref{t:exp3} present the results.

\begin{table}[t]
\centering
\caption{Experiment~3: Final cooperation rates by targeting strategy under 
         compiled influence with grounded agents.}
\label{t:exp3}
\begin{tabular}{lcc}
\hline
\textbf{Targeting strategy} & \textbf{Final cooperation rate} & 
\textbf{Avg cooperation rate} \\
\hline
HUBS      & 0.333 & 0.305 \\
BRIDGES   & 0.333 & 0.305 \\
PERIPHERY & 0.307 & 0.313 \\
RANDOM    & 0.270 & 0.275 \\
\hline
\end{tabular}
\end{table}

The periphery condition's modestly elevated average rate ($0.313$) relative to 
its final value suggests temporary local clustering among low-degree prosumers 
that does not propagate to the wider network. This confirms the 
targeting-over-framing asymmetry identified in~\cite{deCurto2025}: network 
structural position provides mechanistic amplification independent of message 
content, and this persists under grounded agent conditions. \textbf{RQ3} is 
answered affirmatively.

\subsection{Experiment 4: Resistance Sweep (RQ4)}
\label{subsn:exp4}

The fourth experiment sweeps prosumer resistance levels $R \in \{0.1, 0.3, 
0.5, 0.7\}$ and measures the compiled and unstructured cooperation rates at 
each level, testing whether the compilation advantage degrades gracefully or 
collapses beyond a threshold.

Table~\ref{t:exp4} shows the compiled cooperation rate is remarkably stable 
across all resistance levels (range of only $0.007$), while the compilation 
advantage is strictly positive at every level ($+0.057$ to $+0.070$). The 
advantage never vanishes.

\begin{table}[t]
\centering
\caption{Experiment~4: Compilation advantage across resistance levels 
         $R \in \{0.1, 0.3, 0.5, 0.7\}$.}
\label{t:exp4}
\begin{tabular}{lccc}
\hline
\textbf{Resistance $R$} & \textbf{Compiled} & \textbf{Unstructured} & 
\textbf{$\Delta_{\mathrm{comp}}$} \\
\hline
0.1 & 0.330 & 0.263 & +0.067 \\
0.3 & 0.333 & 0.270 & +0.063 \\
0.5 & 0.333 & 0.263 & +0.070 \\
0.7 & 0.327 & 0.270 & +0.057 \\
\hline
\end{tabular}
\end{table}

\begin{figure}[H]
  \centering
  \includegraphics[width=0.8\linewidth]{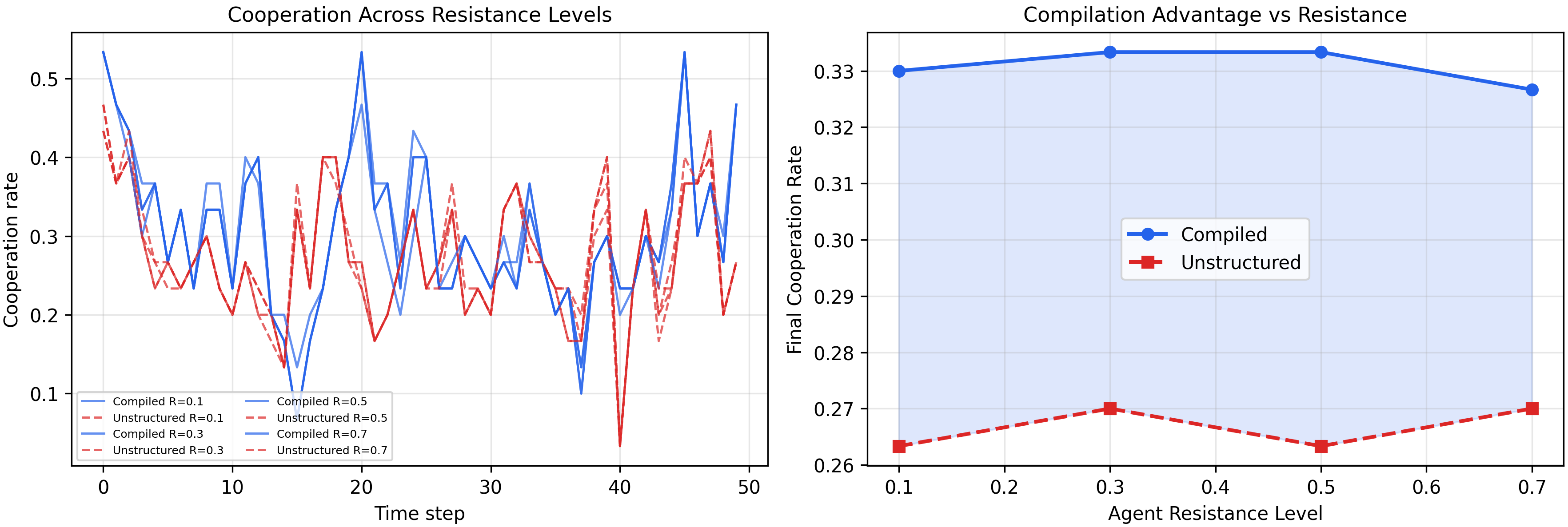}
\caption{Compiled (blue) versus unstructured (red dashed) directives across 
           resistance levels $R \in \{0.1, 0.3, 0.5, 0.7\}$.}
  \label{fgr:resistance_sweep}
\end{figure}

Figure~\ref{fgr:resistance_sweep} confirms the pattern visually. \textbf{RQ4} 
is answered: degradation is graceful, with the compiled condition at $R = 0.7$ 
($0.327$) still substantially exceeding the unstructured condition at $R = 0.1$ 
($0.263$), a compiler design property rather than an artifact of low prosumer 
resistance.

\subsection{Agent-Level Analysis}
\label{subsn:agent}

\begin{figure}[H]
  \centering
  \includegraphics[width=0.8\linewidth]{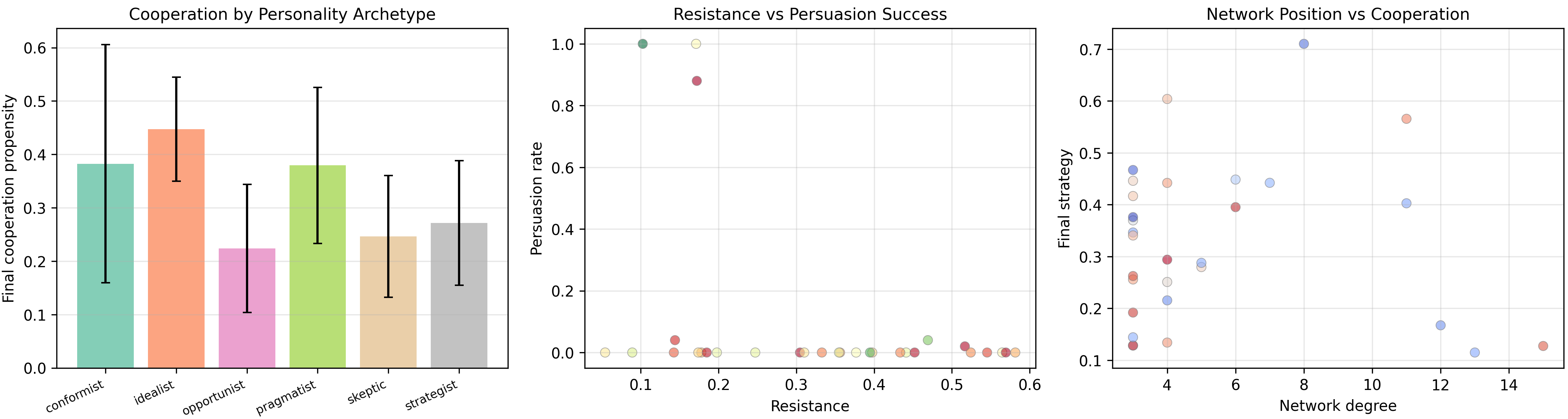}
\caption{Agent-level heterogeneity under compiled influence. Left: final 
           cooperation propensity by archetype. Centre: persuasion rate versus 
           resistance. Right: network degree versus final cooperation strategy.}
  \label{fgr:agent_analysis}
\end{figure}

Figure~\ref{fgr:agent_analysis} decomposes the population-level results into 
agent-level heterogeneity. The left panel confirms that personality archetypes 
produce meaningfully differentiated cooperation propensities under compiled 
influence: idealists achieve the highest final cooperation propensity 
($0.447$), followed by conformists ($0.383$) and pragmatists ($0.379$), while 
opportunists ($0.224$), skeptics ($0.246$), and strategists ($0.272$) cluster 
at lower levels. This differentiation validates the hybrid decision architecture: 
if the RLHF cooperation bias were dominating, all archetypes would converge to 
near-identical high propensities regardless of personality parameters.

The centre panel reveals that persuasion success falls sharply with resistance. 
Agents with $r < 0.20$ (primarily idealists and conformists) are persuaded in 
approximately $85$--$100\%$ of targeted interactions, while agents with 
$r > 0.40$ (skeptics, opportunists) are persuaded in fewer than $5\%$ of 
interactions, and skeptics, never targeted due to their high resistance 
penalty in the compiler's targeting score, register zero persuasion events. 
This confirms that the resistance attenuation in Equation~\ref{e:hybrid} 
operates as intended.

The right panel plots final cooperation strategy against network degree, 
revealing that high-degree hub prosumers tend toward modestly higher final 
cooperation strategies, consistent with the conformist imitation term in 
Equation~\ref{e:base}: well-connected agents receive more cooperative 
neighbourhood signals when hubs are targeted, reinforcing curtailment 
behaviour through social learning~\cite{szabo1998evolutionary,santos2005scale}.

\section{Discussion and Conclusion}
\label{sn:conclusion}

This paper presented a hybrid game-theoretic and LLM narrative evaluation 
architecture for demand-response coordination in smart microgrids, framing 
prosumer curtailment as a repeated Prisoner's Dilemma on a scale-free energy 
community network. The central finding is unambiguous: structured policy 
compilation outperforms unconstrained narrative generation regardless of agent 
sophistication, resistance level, or targeting strategy. The compilation 
advantage of $+0.063$ persists across all resistance levels $R \in [0.1, 0.7]$ 
with no threshold collapse, and hub and bridge targeting consistently dominate 
random and peripheral dissemination, confirming that network topology provides 
mechanistic amplification independent of message content.

The hybrid decision architecture resolves a critical methodological challenge 
for LLM-mediated cooperation research: by separating game-theoretic base 
probability from LLM narrative evaluation, the model prevents RLHF cooperation 
bias from collapsing experimental conditions while preserving grounded 
reasoning as the core influence mechanism. The resulting personality 
differentiation, idealists persuaded in $\sim$85\% of interactions, skeptics 
and opportunists in fewer than 5\%, validates that the architecture produces 
realistic prosumer heterogeneity absent from prior idealized models.

From a smart city energy perspective, these results suggest three practical 
design principles: (1)~demand-response communication systems should employ 
structured directive schemas rather than generic messaging; (2)~dissemination 
should prioritise network hubs and bridges over broad or peripheral outreach; 
and (3)~constitutional governance constraints can filter manipulative or 
inequitable directives with minimal cooperation cost, supporting both grid 
efficiency and prosumer autonomy.

Limitations include small population size ($N = 30$), single network 
realisation per condition, and the use of LLM agents as proxies for human 
prosumers. Validation with human subjects in real energy community settings 
constitutes the primary direction for future work, alongside scaling to 
larger populations and heterogeneous grid topologies.

\section*{Acknowledgments}
This research was supported by the LUXEMBOURG Institute of Science and Technology through the projects ``ADIALab-MAST'' and ``LLMs4EU'' (Grant Agreement No.~101198470) and the BARCELONA Supercomputing Center through the project ``TIFON'' (File number MIG-20232039).

\section*{Data Availability}

The complete experimental data, including all cooperation rate time series, 
agent-level outputs, resistance sweep results, and summary statistics, together 
with all code 
necessary to reproduce the experiments and figures presented in this paper, 
are available at: \url{https://github.com/drdezarza/gice}.

\bibliographystyle{unsrt}
\bibliography{sample}

\end{document}